\documentclass[11pt,prd,showpacs,showkeys]{revtex4}
\usepackage{epsfig}
\usepackage{amsmath}
\usepackage{epsfig}
\usepackage{graphicx}
\begin{document}
\date{February 2nd, 2007}
\title{\begin{flushright}{}\end{flushright}
A model for fermion masses and lepton mixing in $SO(10)\times A_4$}
\author{Stefano Morisi}
\email{stefano.morisi@mi.infn.it}
\affiliation{
Instituto de Fisica Corpuscular (IFIC) - Centro Mixto CSIC-UVEG - Valencia, Spain
}
\author{Marco Picariello}
\email{Marco.Picariello@le.infn.it}
\affiliation{
Dipartimento di Fisica - Universit\`a di Lecce and INFN - Lecce, Italia\\
 and\\
CFTP - Dep. de F\'{\i}sica - Instituto Superior T\'{e}cnico - Lisboa, Portugal
}
\author{Emilio Torrente-Lujan}
\email{torrente@cern.ch}
\affiliation{
Dep. de Fisica, Grupo de Fisica Teorica, Univ. de Murcia - Murcia, Spain.
}
%EndAName

\begin{abstract}
The discrete flavor symmetry $A_4$ explains very well neutrino data at
low energy, but
it seems difficult to extend it  to grand unified models
since in general left-handed and right-handed fields belong to different
$A_4$ representations. Recently it has been
proposed a model where all the fermions equally transform
under $A_4$. We study here a concrete $SO(10)$ realization of such a model providing
small neutrino masses through the seesaw mechanism.
We fit at tree level the charged fermion masses run up to
 the unification scale. Some fermion masses properties
come from the $SO(10)$ symmetry while
lepton mixing angles are consequence of the $A_4$ properties.
Moreover, our model predicts
the absolute value of the neutrino masses, these ones are in the 
range   $m_\nu\simeq 0.005-0.052\ eV$. 

\end{abstract}
\maketitle

\section{Introduction}
The existence of 
 a Grand Unified Theory (GUT) \cite{Pati:1973uk,Georgi:1974sy}
has continued to be 
an attractive idea for physics beyond the Standard Model (SM)
since the 70's. 
Among indications toward GUTs are the phenomenological 
tendency to unify of the gauge couplings,
and the theoretical implicit possibility to explain charge quantization and
 anomaly cancellation.
One of the main features of GUTs is their potentiality to unify
the particle representations and the fundamental parameters
in a hopefully predictive framework. 
There are many gauge group that can accommodate the SM
($SU(5)$, $SU(6)$, $SO(10)$, $E_6$, etc).
Among them $SO(10)$ is the smallest simple Lie group for which a
single anomaly-free irreducible representation (namely the spinor
16 representation) can accommodate the entire SM fermion content
of each generation.

Once we fix the unification group, we  deal with the
flavor physics. The introduction of 
an extra horizontal symmetry 
acting on the fermion families may  further constrain the neutrino
mixing parameters
and hopefully explain large mixing angles.
After the recent neutrino evidence
\cite{SNO}%
%,Krauss:2006qq,Abe:2006fu,SKatm2,SKsolar,GNO,GALLEX,HOMESTAKE,SAGE,KamLAND,CHOOZ,PaloVerde,PaloVerde2,
-\cite{Michael:2006rx}
 we know very well almost all the parameters both
in the quark \cite{Charles:2004jd}
and lepton \cite{Fogli:2005gs}%
%,Aliani:2003ns,Aliani:2002na,Balantekin:2004hi,Oberauer:2004ji,Rodejohann:2006ek,Gonzalez-Garcia:2006wm,Giunti:2006fr,Valle:2006vb,Fogli:2006jk,Bandyopadhyay:2006jn,Bilenky:2006sn,Messier:2006yg,Strumia:2006db,Schwetz:2006dh,Fukugita:2006rm,Chen:2006rk,Robertson:2006pk,Petcov:2006yg,
-\cite{McDonald:2006qf}
sectors.
 We know all the quark and charged lepton masses,
and the value of the difference between the square of the neutrino masses:
$\delta m_{12}^2=m_1^2-m_2^2$ and $\delta m_{23}^2=|m_3^2-m_2^2|$. 
We also know the value of the quark mixing angles and phases, and the two mixing angles
$\theta_{12}$ and $\theta_{23}$ in the lepton sector.
Moreover we have an upper bound for the
$\theta_{13}$ mixing angle in the lepton sector.
All these experimental informations seem to indicate
a discrete flavor symmetry such as 2-3 \cite{23}-\cite{Mohapatra:2004mf}, 
$S3$\cite{Morisi:2005fy}%,S3,Picariello:2006sp
-\cite{Caravaglios:2005gw}, $S4$\cite{S4}-\cite{Hagedorn:2006ug}, $D_3$, 
$D_4$ \cite{D3D4},
$A_4$ \cite{A4}-\cite{King:2006np},
$T^\prime$ \cite{Feruglio:2007uu},
 etc, in the lepton sector. 
In particular, models with $A_4$ flavor symmetry, the case studied
here, very easily give 
the tri-bi-maximal mixing matrix \cite{Harrison:2002er} that fits well
the neutrino data.
Non Abelian discrete symmetries could arise from superstring theory, 
in particular from the compactification of heterotic orbifolds 
\cite{Kobayashi:2006wq}, 
the case for  $A_4$ is reported in \cite{Altarelli:2006kg}. 
Models with $SU(5)\times A_4$ \cite{Ma:2006sk} and 
$SU_L(2)\times SU_R(2)\times SU(4)\times A_4$ \cite{King:2006np}
symmetries have already been studied in literature.
In these previous studies,  fermion 
singlets and $SU_L(2)$ doublets do not equally transform under $A_4$. 
Thus this family symmetry seems not to be not compatible with $SO(10)$ 
models where all the matter fields belong to the same representation.
Only recently it has been proposed a generic 
phenomenological model with $A_4$ \cite{Ma:2006wm}
which is suitable, as we will see in this work, for a SO(10) GUT generalization. 

The purpose of the paper is to construct an explicit $SO(10)\times
A_4$ GUT model and to fit, at tree level,  fermion masses and mixing.
We propose here  a non-SUSY GUT model with a Lagrangian invariant
under $SO(10)\times A_4$. The matter fields are in a ${\bf 16}$,
triplet of $A_4$.
In the Higgs sector, we introduce a ${\bf 10}$, a ${\bf 126}_s$
and three ${\bf 45}$s
singlets of $A_4$, a ${\bf 45}$ and a ${\bf 126}_t$ triplets of $A_4$.
The $A_4$ symmetry is dynamically broken by the vacuum expectation value (vev)
of the Higgs $A_4$-triplets.
The study of the problem of the vacuum alignment in $A_4$
just studied in the context of extra dimensions \cite{Altarelli:2005yp}
and the MSSM \cite{Altarelli:2005yx} is beyond the scope of this work.
The direction of the four vevs of the ${\bf 45}$s in the $SO(10)$
are simply assumed to be $T_{3R}$, $Y$ and two other combinations of them.
The ${\bf 10}$ gives contributions to the Dirac mass matrices
proportional to the identity.
Because of the chosen vev directions and the fact that the ${\bf 45}$s
appear only in a given combination, we get contributions
to $M^u$, $M^d$, $M^l$, but not to $M^\nu_{Dirac}$ from higher 
dimension operators.
The ${\bf 126}$ gives contributions only to the
Majorana neutrino mass matrix. The low energy neutrino mass matrix
is obtained with the see-saw mechanism (for a phenomenological realization
in $A_4$ see \cite{Altarelli:2006ri}).
 
The paper is organized as follows.
In Sec {\bf\ref{sec:so10}} we define the matter and Higgs fields transformations
under the $SO(10)$ and  $A_4$ groups.
In Sec {\bf\ref{sec:Lag}} we write the Lagrangian of our model.
In Sec {\bf\ref{sec:matrices}} we show the relations between the Dirac mass
matrices and the Higgs vevs. 
We show similar relations for the Majorana mass matrix of the neutrinos.
In Sec {\bf\ref{sec:exp}} we write the mixing matrices and masses
as function of the Higgs vevs.
In Sec {\bf\ref{sec:fit}} we show how the experimental data constraint
our model. In Sec {\bf\ref{sec:subexp}}, 
%%with the help of an analytic computations,  %%%  which analitical? 
we perform a numerical analysis of the experimental
data by using a Monte Carlo minimization fit.
In Sec {\bf\ref{sec:subtheo}} we investigate some  predictions of our model.
Sec {\bf\ref{sec:conclusion}} is devoted to conclusions.
We list some relevant, well known, $A_4$ group and representation 
properties in the Appendix {\bf\ref{app:A4}}.

\section{Matter and Higgs fields}
\label{sec:so10}
The smaller spinorial representation of SO(10) is the {\bf 16}
dimensional one. 
All the fermionic matter fields of one family can be accommodated 
within the {\bf 16}, by including the right-handed neutrino.
The Higgs electroweak doublet can be taken in the {\bf 10} as well as one of the 
{\bf 126} representations. For simplicity we assume that the
electroweak doublet Higgs belongs to the 
{\bf 10} representation. Since leptons and quarks mass matrices
cannot be symmetric,
we need to break the $SO(10)$ left-right symmetry at the unification scale.
We perform this job by introducing
sets of fields in the {\bf 45} representation.  The scalar {\bf 45} representations
can get vev in any combination of the extra abelian factors 
$Y$ and $T_{3R}$ directions.
The matter fields and scalar fields transform under $A_4$ as in the table \ref{tab:tab0}
\begin{table}[t]
\begin{center}
\begin{tabular}{|c||c|c|c|c|c|c|c|c|} 
\hline
SO(10) & ~{\bf 16}~ & ~{\bf 10}~ & ~{\bf 45$_{T_{3R}}~$} & ${\bf 45}_Y$& ${\bf 45}_C$& ${\bf 45}_D$&
 {\bf 126}$_s$&{\bf126}$_t$\\
\hline
$A_4$  & 3  &1   &1       &  1  & 3&1  &1&3 \\
\hline
\end{tabular}
\caption{Matter and Higgs field representations}\label{tab:tab0}
\end{center}
\end{table}
where the index of the ${\bf 45}$s reefers to the vevs direction.
$C$ and $D$ are linear combinations of $Y$ and $T_{3R}$.
We will determine these combinations latter, by using the experimental
constraints.

\section{The Lagrangian}\label{sec:Lag}
Let us write our Lagrangian, it reads
\begin{eqnarray}\label{eq:Lag}
L_\text{Y}&=&h_0^{ij} {\bf16}^i~{\bf10~16}^j
  +h_0^{\prime\,ij}{\bf16}^i~{\bf 10}~{\bf45}_{T_{3R}}~{\bf45}_Y~{\bf16}^j
  +h^{ijk}{\bf16}^i~{\bf 10}~{\bf45}_{T_{3R}}~{\bf45}_Y~
  {\bf45}^j_C~{\bf45}_D~{\bf16}^k
\\\nonumber&&
 +  \sigma^{il}{\bf16}^i~{\bf45}_{T_{3R}}~{\bf126}_s~{\bf45}_{T_{3R}}~{\bf16}^j
 +\lambda^{ijk}{\bf16}^i~{\bf45}_{T_{3R}}~{\bf126}^j_t~{\bf45}_{T_{3R}}~{\bf16}^k
\\&\equiv& L_\text{Dirac}+ L_\text{Majo}
\end{eqnarray}
where the indices $\{i,j,k,l\}$ are  $A_4$ indeces
 and the sum over the gauge indices is understood.
As shown in \cite{Anderson:1993fe} any Lagrangian of the form in eq. (\ref{eq:Lag})
can be easily obtained from a renormalizable Lagrangian, by including a set
of heavy spinor fields, with the inclusion of an $U(1)$ charge and/or
 super-symmetry.
We reserve to a further investigation the question
 of how general is our Lagrangian, and how it can be obtained in
a renormalizable theory.

As we will better clarify in the Appendix {\bf\ref{app:A4}},
in the second and in the last terms of eq.(\ref{eq:Lag}) there are two ways
of contracting the three $A_4$ indices in an invariant way.
We have to choose to which representation of $A_4$ the {\bf 10} scalar field belongs.
Because we want only one Higgs, we excluded the triplet possibility
but we have still three possibilities that correspond to 
how the {\bf 10} transforms with
respect to $A_4$: as {\bf 1}, ${\bf 1'}$, ${\bf 1''}$.
The  fermion mass matrices $M_f$ (with $f=u,d,l,\nu$) coming from the first term in $L_Y$
will be, respectively:
\begin{equation}
\left(
\begin{array}{ccc}
1&0&0\\
0&1&0\\
0&0&1
\end{array}
\right),~
\left(
\begin{array}{ccc}
1&0&0\\
0&\omega^2&0\\
0&0&\omega
\end{array}
\right),~
\left(
\begin{array}{ccc}
1&0&0\\
0&\omega&0\\
0&0&\omega^2
\end{array}
\right).
\end{equation}
In any case we have three degenerate eigenvalues, namely $m_u=m_c=m_t$, 
that are corrected by the additional terms in eq.(\ref{eq:Lag}).
Let us assume that the $A_4$ triplets ${\bf45}_C$ and {\bf126}$_t$ get 
vevs, respectively, in the
following directions of $A_4$ 
\begin{eqnarray}
\langle {\bf 45}_C \rangle = v_{{\bf 45}_C}(1,1,1),&\quad\quad\quad&
~~\langle {\bf126}_t \rangle = v_{{\bf 126}_t}(1,0,0)
\end{eqnarray}
where the $SO(10)$ indices are understood on both left and right sides.
After symmetry breaking, 
once the Higgs acquire vevs, the quadratic part for the fermions
of the Lagrangian in eq.(\ref{eq:Lag}) can be rewritten as
\begin{subequations}
\begin{eqnarray}
L_\text{Dirac}&=&
h_0~({\bf16}_1{\bf16}_1+{\bf16}_2{\bf16}_2+{\bf16}_3{\bf16}_3)~v_{\bf 10}+
\\&+&
h_0^\prime~({\bf16}_1{\bf16}_1^\prime+{\bf16}_2{\bf16}_2^\prime+{\bf16}_3{\bf16}_3^\prime)~v_{\bf 10}+\label{eq:diag}\\
&+&
h_1~({\bf16}_1{\bf16}_2^{\prime\prime}+{\bf16}_2{\bf16}_3^{\prime\prime}+{\bf16}_3{\bf16}_1^{\prime\prime})~v_{\bf10}\label{eq:h1}+\\
&+&
h_2~({\bf16}_1{\bf16}_3^{\prime\prime}+{\bf16}_2{\bf16}_1^{\prime\prime}+{\bf16}_3{\bf16}_2^{\prime\prime})~v_{\bf 10}\label{eq:h2}\\
L_\text{Majo}&=&\sigma({\bf16}_1^{\prime\prime\prime}{\bf16}_1^{\prime\prime\prime}+{\bf16}_2^{\prime\prime\prime}{\bf16}_2^{\prime\prime\prime}+
{\bf16}_3^{\prime\prime\prime}{\bf16}_3^{\prime\prime\prime})~v_{{\bf126}_s}
+\lambda{\bf16}_2^{\prime\prime\prime}{\bf16}_3^{\prime\prime\prime}~v_{{\bf126}_t}
\end{eqnarray}
\end{subequations}
where 
\begin{eqnarray}
{\bf16}_i^{\prime\prime}\equiv v_{{\bf45}_{T_{3R}}}~v_{{\bf45}_Y}~v_{{\bf45}_C}~v_{{\bf45}_D}~{\bf16}_i
&\quad\quad&
{\bf16}_i^{\prime\prime\prime}\equiv v_{{\bf45}_{T_{3R}}}~{\bf16}_i\\
{\bf16}_i^{\prime}\equiv v_{{\bf45}_{T_{3R}}}~v_{{\bf45}_Y}~{\bf16}_i
&\quad\quad&
\text{with}~~i=1,2,3\nonumber
\end{eqnarray}
 We obtain the following expression by absorbing 
  the vevs of the ${\bf 45}$s into the coupling constants
\begin{subequations}
\begin{eqnarray}
{\bf 16^{\prime}}&=&
\left(
{{x_q}_L}~q,
~{{x_u}_R}~u^c,
~{{x_d}_R}~d^c,
~{{x_l}_L}~l,
~{{x_e}_R}~e^c,
~{{x_\nu}_R}~\nu_R
\right)^T,
\\
{\bf 16^{\prime\prime}}&=&
\left(
{{x'_q}_L}~q,
~{{x'_u}_R}~u^c,
~{{x'_d}_R}~d^c,
~{{x'_l}_L}~l,
~{{x'_e}_R}~e^c,
~{{x'_\nu}_R}~\nu_R
\right)^T,
\\
{\bf 16}^{\prime\prime\prime}&=&
\left(
{{x''_q}_L}~q,
~{{x''_u}_R}~u^c,
~{{x''_d}_R}~d^c,
~{{x''_l}_L}~l,
~{{x''_e}_R}~e^c,
~{{x''_\nu}_R}~\nu_R\right)^T
\end{eqnarray}
\end{subequations}
where ${x_f}_{L,R}$, ${{x'_f}_{L,R}}$, and ${{x''_f}_{L,R}}$ are the quantum numbers respectively of the product of the charges ${T_{3R}}$ with $Y$, of
the product of the charges ${T_{3R}}$, $Y$, $C$, and $D$, and of the charge ${T_{3R}}$ reported in Table (\ref{tab:tab1}) \cite{Anderson:1993fe}.
\begin{table}[t]
\begin{center}
\begin{tabular}{|c|c|c|c|c|}
\hline
\quad\ \quad\ & \quad $X$ \quad\ & \quad $Y$ \quad\ & $B-L$ &\ \ $T_{3R}$\ \ \\
\hline
$q$    & 1 & 1/3& 1&0   \\
$u^c$  & 1 &-4/3&-1&1/2 \\
$d^c$  & -3& 2/3&-1&-1/2\\
$l$    & -3&  -1&-3&   0\\
$e^c$  & 1 &   2& 3&-1/2\\
$\nu^c$& 5 &   0& 3&1/2 \\
\hline
\end{tabular}
\caption{Quantum numbers for the low energy matter fields.}\label{tab:tab1}
\end{center}
\end{table}

\section{From vevs to mass matrices}\label{sec:matrices}
From Table {\ref{tab:tab1} we observe that
${{x'_\nu}_R}=0$ (because $Y$ of the right handed neutrino is zero)
and
${{x'_l}_L}=0$ (because $T_{3R}$ of the lepton doublet is zero).
These two conditions imply that the terms
${\bf 16_i~16_j^{\prime\prime}}v_{\bf 10}$ in  the Lagrangian $L_{Dirac}$
do not
contribute to the Dirac neutrino mass term. Therefore, once the {\bf 45}s
get a vev, from eq.(\ref{eq:diag}) we have
that the Dirac neutrino mass matrix $M^\nu_{Dirac}$ is proportional to the identity.
\begin{equation}\label{eq:Dirac}
M^\nu_{Dirac}~=h_0~v^u~{\bf I}~,
\end{equation} 
where ${\bf I}$ is the identity matrix and $v^u$ is the vev of the 
up component of the ${\bf 10}$
\footnote{
Because the presence of the product
${\bf 45}_Y\,{\bf 45}_{T_{3R}}$ in the $h_0^\prime$ term,
the only contribution to $M^\nu_{Dirac}$ comes from $h_0$.}.
The fact that the $M^\nu_{Dirac}$ is proportional to the identity
 will be important
in order to realize the see-saw mechanism and not spoiling the 
main consequence 
 of the $A_4$ symmetry: the explanation of the appearence of 
a tri-bi-maximal mixing matrix in the lepton sector.
With the conventions
 ${x_u}_L={x_d}_L\equiv{x_q}_L$, ${x_e}_L={x_\nu}_L\equiv{x_l}_L$, and $v^e=v^d$, 
the interactions $h_1~{\bf 16_1~16_2^{\prime\prime}}$
 and $h_2~{\bf 16_2~16_1^{\prime\prime}}$
in eqs.(\ref{eq:h1}) and (\ref{eq:h2}) give the following mass terms
\begin{eqnarray}
h_1 v^f~(x'_{fL}~\overline{\psi}_{L1}~\psi_{R2}
     + x'_{fR}~\overline{\psi}_{L2}~\psi_{R1})+
h_2 v^f~(x'_{fL}~\overline{\psi}_{L2}~\psi_{R1}
   + x'_{fR}~\overline{\psi}_{L1}~\psi_{R2})~{+h.c.}
\end{eqnarray}
namely,
$$
v^f\left(
\begin{array}{cc}
0&h_1~x'_{fL}+h_2~x'_{fR}\\
h_1~x'_{fR}+h_2~x'_{fL}&0
\end{array}
\right)_{12}
$$
and so on for the other interactions (in the flavor planes 31 and 23). 
If we introduce
\begin{equation}\label{eq:AB}
A^f= (h_1~ x'_{fL}+h_2~x'_{fR})
\quad\mbox{and}\quad~~B^f=(h_1~x'_{fR}+h_2~x'_{fL})
\end{equation}
the full contribution to the Dirac mass matrices,
coming from the operators proportional to the {\bf 45}
representations, is
\begin{equation}
v^f \left(
\begin{array}{ccc}
0&A^f&B^f\\
B^f&0&A^f\\
A^f&B^f&0
\end{array}
\right)
\end{equation}
The charged fermion mass matrices are then 
\begin{equation}\label{eq:mferm}
M^{u}=v^u\left(
\begin{array}{ccc}
h_0^u&A^u&B^u\\
B^u&h_0^u&A^u\\
A^u&B^u&h_0^u
\end{array}
\right);~~
M^{d}=v^d\left(
\begin{array}{ccc}
h_0^d&A^{d,l}&B^{d,l}\\
B^{d,l}&h_0^d&A^{d,l}\\
A^{d,l}&B^{d,l}&h_0^d
\end{array}
\right);~~
M^{l}=v^d\left(
\begin{array}{ccc}
h_0^l&A^{d,l}&B^{d,l}\\
B^{d,l}&h_0^l&A^{d,l}\\
A^{d,l}&B^{d,l}&h_0^l
\end{array}
\right)
\end{equation}
where $v^u$ and $v^d$ are the vevs of the up and down components
of the {\bf 10}, while the $A$ and $B$ coefficients are defined in 
eq.(\ref{eq:AB}).
The $h_0^f$ are defined by the combinations of $h_0$ and $h_0^\prime$
with the weight corresponding to the charge
$x_{fR}$:
\begin{subequations}
\begin{eqnarray}
h_0^u&=& h_0 + x_{uR}\ h_0^\prime,\\
h_0^d&=& h_0 + x_{dR}\ h_0^\prime,\\
h_0^l&=& h_0 + x_{eR}\ h_0^\prime.
\end{eqnarray}
\end{subequations}
We observe that the general form of the mass matrices
$M^{u,d,l}$, are of the same type of the one reported 
%by E.Ma 
in Ref. 
\cite{Ma:2006vq} (see eq.20).
Moreover the Majorana mass matrix for the right-handed neutrino is given by
\begin{equation}
M_R=\left(
\begin{array}{ccc}
a&0&0\\
0&a&b\\
0&b&a
\end{array}
\right)
\end{equation}
where $a=\sigma~ v_{{\bf 126}_s}$ and $b=\lambda~ v_{{\bf 126}_t}$. The 
Dirac neutrino mass matrix has been previously 
 given in  eq.~(\ref{eq:Dirac}). 

\section{Masses and Mixing}\label{sec:exp}

It has  been recently shown in \cite{Ma:2006vq} that,
if the Dirac mass matrices are given by eq.(\ref{eq:mferm}),
the charged fermion mass matrices are diagonalized by 
\begin{equation}\label{eq:U}
U=\frac{1}{\sqrt{3}}\left(
\begin{array}{ccc}
1&1&1\\
1&\omega&\omega^2\\
1&\omega^2&\omega
\end{array}
\right)
\end{equation}
and then we have
\begin{eqnarray}\label{eq:masses}
M^{f}&=&U~\left(
\begin{array}{ccc}
(h_0^f +A^f+B^f)~v^f &0&0\\
0&(h_0^f +\omega A^f+\omega^2 B^f)~v^f&0\\
0&0&(h_0^f+\omega B^f+\omega^2 A^f)~v^f
\end{array}
\right)~U^\dagger
\end{eqnarray}
where $f=u,d,l$, $v^l=v^d$, and $h^f_0,~A^f$ and $B^f$ 
are complex parameters.

From the Lagrangian in eq. (\ref{eq:Lag}),
the light neutrino mass matrix comes from a type-I seesaw 
mechanism as below 
\begin{eqnarray} 
M^\nu&=&M^{\nu}_{Dirac}\frac{1}{M_{R}}\left(M^{\nu}_{Dirac}\right)^T
\end{eqnarray}
where the Dirac neutrino mass matrix $M^{\nu}_{Dirac}$ 
is proportional to the identity (see eq.~\ref{eq:Dirac}), 
while $M_{R}$ is the right handed Majorana neutrino matrix. 
We observe that our Lagrangian does not give the left handed
$M_L$ Majorana neutrino
 mass matrix since we have introduced the $T_{3R}$ fields. 
In the basis where the charged leptons are diagonal, the mass matrix of the 
low energy neutrino $\overline{M_\nu}$ is given by
\begin{eqnarray}
\overline{M_\nu}=U^T M_\nu U = M^\nu_{Dirac} \frac{1}{U^\dagger M_R U^\star} \left(M^\nu_{Dirac}\right)^T
\end{eqnarray}
where we used the fact that $M^\nu_{Dirac}$ is proportional to the identity.
We have
\begin{equation}
U^\dagger M_R U^\star=\left(
\begin{array}{ccc}
a+2b/3&-b/3&-b/3\\
-b/3&2b/3&a-b/3\\
-b/3&a-b/3&2b/3
\end{array}
\right)
\end{equation}
and it is diagonalized by a tri-bi-maximal mixing matrix.
Consequently $\overline{M_\nu}$ is diagonalized by the same 
tri-bi-maximal mixing
matrix too. 
The eigenvalues of $\overline{M^\nu}$ are
\begin{subequations}
\begin{eqnarray}
m_1&=&\frac{(h_0 v^{u})^2}{a+b},\\
m_2&=&\frac{(h_0 v^{u})^2}{a},\\
m_3&=&\frac{(h_0 v^{u})^2}{b-a}.
\end{eqnarray}
\end{subequations}

\section{Numerical fitting and model predictions}\label{sec:fit}
In the following subsection {\bf\ref{sec:subexp}}, we analyze
how to translate
% well the charged fermion mass matrices in eq. (\ref{eq:masses})
%are compatible with the experimental data. 
%In particular we will
all the informations from the experimental data
into constraints for the parameters of our theory.
Then, in the subsection {\bf\ref{sec:subtheo}} 
%we follow a different point of view,
we will show how well 
the charged fermion mass matrices in eq. (\ref{eq:masses})
can be fitted. We also include some theoretical predictions
of our model about the absolute neutrino masses.

\subsection{Experimental constraints}\label{sec:subexp}
From eq.(\ref{eq:masses}) we have that the tree mass eigenvalues
for the charged fermions are of the form
\begin{subequations}\label{eq:system}
\begin{eqnarray}
(h_0^f+A^f+B^f)~v^f &=&m_1^f,\\
(h_0^f+\omega A^f+\omega^2B^f)~v^f &=&m_2^f\label{hab1},\\
(h_0^f+\omega^2 A^f+\omega B^f)~v^f &=&m_3^f.
\end{eqnarray}
\end{subequations}
where the masses $m_i^f$ are in general complex and their phases are 
unphysical. 
The parameters $h_0^f$, $A^f$, and $B^f$ are complex.
The $v^f$ are the vevs of the scalar
Higgs doublets in the {\bf 10} and $v^l~=~v^d$. 
The most general solution of the system in eq. (\ref{eq:system}) is
\begin{subequations}\label{eq:sol}
\begin{eqnarray}
h_0^f&=& \frac{1}{v^f}\frac{m_1^f+m_2^f+m_3^f}{3}\\
A^f  &=& \frac{1}{v^f}\frac{m_2^f\omega^2+m_1^f + m_3^f \omega }{3}\\
B^f  &=& \frac{1}{v^f}\frac{m_3^f\omega^2+m_1^f + m_2^f \omega }{3}
\end{eqnarray}
\end{subequations}
The numerical values of $h_0^f$, $A^f$ and $B^f$ in eq.~(\ref{eq:sol})
are then fixed, up to phases, by the fermion masses.
The absolute value of $h_0^f$ can be written as
\begin{eqnarray}\label{eq:h0f}
|h_0^f|^2&=&\left(\frac{1}{3 v^f}\right)^2\Big[
\Big(m_1^f + m_2^f + m_3^f\Big)^2\nonumber\\
&&-
2 \Big( m_1^f\ m_3^f(1-\cos\phi_1) + m_1^f\ m_2^f(1-\cos(\phi_1 - \phi_2)) + 
    m_2^f\ m_3^f(1-\cos\phi_2)\Big)\Big]
\end{eqnarray}
where $\phi_1$ and $\phi_2$ are the relative phases between
 $m_1$ and $m_3$ and between $m_2$ and $m_3$ respectively.
From eq.(\ref{eq:h0f}) and by assuming that $m_3>m_1+m_2$, we obtain 
\begin{subequations}
\begin{eqnarray}
\frac{1}{3 v^f} \Big( m_1^f + m_2^f + m_3^f\Big)
 \geq |h_0^f| \geq
\frac{1}{3 v^f} \Big( m_3^f - m_1^f - m_2^f\Big)
\end{eqnarray}
In the same manner we get
\begin{eqnarray}
\frac{1}{3 v^f} \Big( m_1^f + m_2^f + m_3^f\Big)
 \geq |A^f| \geq
\frac{1}{3 v^f} \Big( m_3^f - m_1^f - m_2^f\Big)
\\
\frac{1}{3 v^f} \Big( m_1^f + m_2^f + m_3^f\Big)
 \geq |B^f| \geq
\frac{1}{3 v^f} \Big( m_3^f - m_1^f - m_2^f\Big)
\end{eqnarray}
\end{subequations}
Under the condition that $m_3 \gg m_1, m_2$,
 the phases among $h_0^f$, $A^f$ and $B^f$ are strongly constrained
by the last equation in eq.(\ref{eq:system}). From the solutions
in eq.(\ref{eq:sol}) we get
\begin{eqnarray}
\frac{A^f}{h_0}\simeq\omega &\quad\mbox{and}\quad&
\frac{B^f}{h_0}\simeq\omega^2
\end{eqnarray}
%
%In case of real masses $m_i^f$ we obtain
%\begin{subequations}
%\begin{eqnarray}
%\tilde{A}^f&\equiv& \frac{a^f+i~b^f}{v^f}\\
%\tilde{B}^f&\equiv& \frac{a^f-i~b^f}{v^f}
%\end{eqnarray}
%\end{subequations}
%where we have used $\omega=-1/2+\sqrt{3}/2~i$ and defined
%\begin{equation}\label{eq:hab3}
%a^f=\frac{2 m_1^f-m_2^f-m_3^f}{6};~~b^f=\frac{m_3^f-m_2^f}{2\sqrt{3}}.
%\end{equation}
%and there is no indetermination.
From the solution in eq.(\ref{eq:sol}) and by using the
definitions of $A^f$, $B^f$ in eq.(\ref{eq:AB}) we obtain
\begin{eqnarray}\label{eq:xfromm}
{x'_+}^f = \frac{1}{3 v^f}\frac{m_3^f+m_2^f-2m_1^f}{h_1+h_2}
&\quad\mbox{and}\quad&
{x'_-}^f = \frac{i}{\sqrt{3} v^f}\frac{m_3^f-m_2^f}{h_1-h_2}
\end{eqnarray}
where we have introduced the notation ${x'_\pm}^f\equiv {x'_L}^f\pm {x'_R}^f$.
In eq. (\ref{eq:xfromm}) we must remember that each mass 
includes an undetermined phase.
We notice that the ratios ${x'_+}^f/{x'_+}^{f^\prime}$ and ${x'_-}^f/{x'_-}^{f^\prime}$ do not depend on $h_i$, then
they are experimentally determined (up to the undetermined phases).
In fact we have
\begin{subequations}\label{eq:xpxm}
\begin{eqnarray}
\frac{{x'_+}^u}{{x'_+}^d}&=&\frac{v^d}{v^u}\ \frac{m_t+m_c-2m_u}{m_b+m_s-2m_d},
\\
\frac{{x'_-}^u}{{x'_-}^d}&=&\frac{v^d}{v^u}\ \frac{m_t-m_c}{m_b-m_s},
\\
\frac{{x'_+}^u}{{x'_+}^e}&=&\frac{v^d}{v^u}\ \frac{m_t+m_c-2m_u}{m_\mu+m_\tau-2m_e},
\\
\frac{{x'_-}^u}{{x'_-}^e}&=&\frac{v^d}{v^u}\ \frac{m_t-m_c}{m_\tau-m_\mu}.
\end{eqnarray}
\end{subequations}
By using the masses run up to the  $2 \cdot 10^{16}$GeV scale in the
(non SUSY) Standard Model given in Table \ref{tab:run}, we performed 
a Monte Carlo analysis of eq. (\ref{eq:xpxm}). 
For the masses we took two sided Gaussian distributions with
central values and standard deviations taken from Table \ref{tab:run}.
For the unknown phases we took flat random distributions in the
interval $[0, 2\pi]$.
Our results can be summarized as
\begin{subequations}\label{eq:ratio}
\begin{eqnarray}
\frac{{x'_+}^u}{{x'_+}^d}=0.972^{+0.073}_{-0.013}
&\quad\quad\quad&
\frac{{x'_-}^u}{{x'_-}^d}=1.034^{+0.007}_{-0.072}\\
\frac{{x'_+}^u}{{x'_+}^e}=0.573^{+0.079}_{-0.011}
&\quad\quad\quad&
\frac{{x'_-}^u}{{x'_-}^e}=0.640^{+0.011}_{-0.077}\\
\frac{{x'_+}^d}{{x'_+}^e}=0.590^{+0.085}_{-0.048}
&\quad\quad\quad&
\frac{{x'_-}^d}{{x'_-}^e}=0.619^{+0.054}_{-0.075}
\end{eqnarray}
\end{subequations}
Notice that, if we neglect the undetermined phases in the masses, we
get similar central values  but wrong errors in the constraints. For 
example we would obtain in such a case:
\begin{eqnarray}
\frac{{x'_+}^u}{{x'_+}^d}=0.972\pm0.005,
&\quad\quad\quad&
\frac{{x'_-}^u}{{x'_-}^d}=1.034\pm0.006.
\end{eqnarray}

\begin{table}[t]
\begin{center}
{\begin{tabular}{lrrr}
\hline
\hspace{4.cm}&$m_u$(MeV)&   0.8351$^{+0.1636}_{-0.1700}$&\hspace{4.cm}\\
%&&&\\
&$m_c$(MeV)& 242.6476$^{+23.5536}_{-24.7026}$&\\
%&&&\\
&$m_t$(GeV)&  75.4348$^{+9.9647}_{-8.5401}$&\\
%&&&\\
&$m_d$(MeV)&   1.7372$^{+0.4846}_{-0.2636}$&\\
%&&&\\
&$m_s$(MeV)&  34.5971$^{+4.8857}_{-5.1971}$&\\
%&&&\\
&$m_b$(GeV)&   0.9574$^{+0.0037}_{-0.0169}$&\\
&$m_e$(MeV)&   0.4414$^{+0.0001}_{-0.0001}$&\\
&$m_\mu$(MeV)&93.1431$^{+0.0136}_{-0.0101}$&\\
&$m_\tau$(GeV)&1.5834$^{+10.4664}_{-13.6336}$&\\
\hline
\end{tabular}}
\caption{Quark masses run at the  $2\cdot 10^{16}Gev$ scale in non SUSY standard model (see Ref.~\cite{Das:2000uk}).}
\label{tab:run}
\end{center}
\end{table}
 
\subsection{The theoretical prediction}\label{sec:subtheo}
In our model we are able to fit all the masses of quarks and leptons.
Moreover we obtain, thanks to the $A_4$ structure of the model, a
tri-bi-maximal lepton mixing matrix.
Let us investigate the fermion masses. As obtained in the
previous section, the quantities to be fitted are the ratios in 
eq. (\ref{eq:ratio}). The theoretical result for the ratios
$x_+^f/x_+^{f^\prime}$ and $x_-^f/x_-^{f^\prime}$
are determined from Table \ref{tab:tab0} and the
definitions of $x_\pm^f$.
By using for example the direction
$C=(28 X - 249 Y)$ and $D=(238 X - 9 Y)$ we get
\begin{subequations}
\begin{eqnarray}
\frac{{x'_+}^u}{{x'_+}^{d}}=1
&\quad\mbox{and}\quad&
\frac{{x'_-}^u}{{x'_-}^{d}}=1\,;
\\
\frac{{x'_+}^u}{{x'_+}^{e}}=\frac{300}{517}
&\quad\mbox{and}\quad&
\frac{{x'_-}^u}{{x'_-}^{e}}=\frac{300}{517}\,.
\end{eqnarray}
\end{subequations}
in good agreement with the experimental values in eq. (\ref{eq:ratio}).
The absolute neutrino mass scale is fixed, because the
presence of, essentially, only two free parameters, $a$ and $b$, in the neutrino sector.
If we impose the experimental constraints on
$\delta m_{12}^2=7.92\left(1\pm0.09\right)\times 10^{-5} eV^2$ and 
$|\delta m_{13}^2|=2.4\left(1^{+0.21}_{-0.26}\right)\times10^{-3}eV^2$
we get the following neutrino masses:
\begin{eqnarray}
&& m_1=0.052\pm0.005\ eV\,,\quad\quad
m_2=0.052\pm0.005\ eV\,,\quad\quad
m_3=0.017\pm0.002\ eV
\\
&& m_1=0.0051\pm0.0005\ eV\,,\quad\quad
m_2=0.0102\pm0.0005\ eV\,,\quad\quad
m_3=0.049\pm0.004\ eV
\end{eqnarray}
where the first results correspond to a  Inverted Hierarchy case,
while the second ones would correspond to the Normal Hierarchy. 
It is easily possible to see that these 
masses are independent of the phases of the complex parameters $a,b$

\section{Conclusions}\label{sec:conclusion}
Neutrino data at low energy are well explained by a $A_4$
symmetry, nevertheless it is 
difficult to include this symmetry in grand unified theories.
In this paper we investigate the possibility to construct an explicit
model with a Lagrangian invariant under $SO(10)\times A_4$.
We assumed that the matter fields are in a ${\bf 16}$ dimensional 
$SO(10)$ representation, triplet of $A_4$.
In the Higgs sector, we introduced a ${\bf 10}$, a ${\bf 126}$ and three 
${\bf 45}$s
singlets of $A_4$, a ${\bf 45}$ and a ${\bf 126}$ triplets of $A_4$.
The $A_4$ symmetry is dynamically broken by the vevs of the Higgs
$A_4$-triplets.
The direction of the vevs of the ${\bf 45}$s in the $SO(10)$
are assumed to be $T_{3R}$, $Y$ and two other combinations
of them, $C$ and $D$.
The Lagrangian contains three terms with the ${\bf 10}$
that give contributions to the Dirac mass matrices,
and two terms with the ${\bf 126}$s that determine the
Majorana neutrino mass matrix.
The first two terms containing the ${\bf 10}$
give a contribution to the Dirac mass matrices
which is proportional to the identity
(the second term is used to avoid the $\tau$ bottom unification).
The third term, because of the fact
that the ${\bf 45}$s appear only in the given combination,
provides  contributions to $M^u$, $M^d$, $M^l$, 
but not to $M^\nu_{Dirac}$.
For these reasons $M^\nu_{Dirac}$ results to be proportional to the
identity.
Finally the ${\bf 126}$ terms give contribution only to the
right handed neutrino Majorana mass matrix $M_R$.
The low energy neutrino mass matrix is then obtained with the
see-saw mechanism.

The mixing angle structure  of the charged fermion mass matrices are fixed
by the $A_4$ structure of the model. They are diagonalized
by the mixing matrix in eq. (\ref{eq:U}).
The $A_4$ direction of the vev of the triplet
${\bf 126}$ implies a particular form for
$M_R$. This particular form of $M_R$ and the fact that $M^\nu_{Dirac}$
is proportional to the identity, imply that the low energy
neutrino mass matrix, in the base with diagonal charged lepton,
is diagonalized by the tri-bi-maximal mixing matrix.

We show that at tree level our model fits with great
precision (within one standard deviation) the values
of the fermion masses, run at $2\cdot 10^{16}$GeV scale
in the (non SUSY) Standard Model, if particular
 directions of the
vevs of the ${\bf 45}_C$ and ${\bf 45}_D$ are assumed 
($C=(28 X - 249 Y)$ and $D=(238 X - 9 Y)$).

One important consequence of the structure of this model is 
the prediction of an absolute scale for low mass neutrinos.
We predict the absolute scale of the neutrino mass
to be close to $\sim 0.05\ eV$. Normal or inverted hierarchies 
are allowed by the model.

In the model presented here, 
 both up and down sector are diagonalized by the same mixing
matrix. For this reason the resulting quark mixing matrix, the 
CKM matrix is proportional to the identity, in agreement with 
evidence only at first order.
The explanation of the correct CKM matrix is beyond the scope of 
this work. However a  deeper study of  radiative corrections
to the potential could posibly shed light on the right 
 CKM structure.

\subsection*{Acknowledgments}
S.M. thanks Guido Altarelli for the encouragement and for clarifications
about $A_4$ and Francesco Caravaglios
for the clarifications about unified theories. 
We acknowledge a MEC-INFN grant, Fundacion Seneca
(Comunidad Autonoma de Murcia) grant, a
CYCIT-Ministerio of Educacion (Spain) grant and to 
the Funda\c{c}\~{a}o para a Ci\^{e}ncia e a Tecnologia for the grant SFRH/BPD/25019/2005.
We acknowledge the kind hospitality of the Dipartimento di Fisica
dell'Universit\`a degli Studi di Milano where part of this work was done.

\appendix
\section{The $A_4$ properties}\label{app:A4}
The group 
$A_4$ is the finite group of the even permutations of four object and 
contains 12 elements. 
Every finite group can be generated by a
 subset  of elements, called generators. 
A set of elements is independent
if none of them can be expressed in terms of the other. The group
 $A_4$ has two independent generators
denoted as $S$ and $T$, which can be chosen to 
 to verify the following defining relations:
$$S^2=T^3=(ST)^3=I.$$
There are four irreducible
representations for the  $A_4$ group: denoted as $1,1',1''$ and the
$3$. In each of these representations the generators are explicitly
written
as follows
\begin{eqnarray}
1  &:&~ S=1,~ T=1, \nonumber\\
1' &:&~ S=1,~ T=\omega, \nonumber\\
1''&:&~ S=1,~ T=\omega^2, \nonumber\\
3  &:&~ 
S=\left(\begin{array}{ccc}
1&0&0\\
0&-1&0\\
0&0&-1
\end{array}\right),~ 
T= \left(\begin{array}{ccc}
0&1&0\\
0&0&1\\
1&0&0
\end{array}\right),
\end{eqnarray} 
where $\omega=e^{2\pi i/3}$ and then $\omega^3=1$ and $1+\omega+\omega^2=0$.
If $a=(a_1,a_2,a_3)$ is a triplet, then 
the action of the $S$ and $T$ operators is 
$S a=(a_1,-a_2,-a_3)$ and $T a = (a_2,a_3,a_1)$.
If $b$ is another analogous $A_4$ triplet, their tensorial 
product decomposes in irreducible representations as 
$$3\times 3=1+1'+1''+3+3.$$
In order to explicitly construct a  singlet from these quantities 
we first impose the invariance under $S$, the most generic term 
will be
$$ x~a_1b_1+y~a_2b_2+z~a_3b_3+t~a_2b_3+r~a_3b_2$$
where $x,y,z,r$ and $t$ are parameters. If we impose also the invariance under T, we have 
that the above term transforms like a 1 single, if and only if $x=y=z$ and $r=t=0$. Then we have 
\begin{eqnarray*}
1&=&(ab)=(a_1~b_1+a_2~b_2+a_3~b_3)
\end{eqnarray*}
Similarly one can check that
\begin{eqnarray*}
1'&=&(ab)'=(a_1~b_1+\omega^2a_2~b_2+\omega a_3~b_3),\\
1''&=&(ab)''=(a_1~b_1+\omega a_2~b_2+\omega^2 a_3~b_3).
%3&=&(a_2 b_3,~a_3b_1,~a_1 b_2)\\
%3&=&(a_3 b_2,~a_1b_3,~a_2 b_1)
\end{eqnarray*}

Let us go now to construct the triplet.
By imposing  $S$ invariance, the most generic triplet in the
product of  $a$ and $b$ is 
$$(x~a_1b_1+y~a_2b_2+z~a_3b_3+t~a_2b_3+r a_3 b_2,\tilde{x}~a_1b_2+
\tilde{y}~a_1b_3+\tilde{z}~a_2b_1+\tilde{t}~a_3b_1,...)$$
applying $T$ we have
$$(x~a_2b_2+y~a_3b_3+z~a_1b_1+t~a_3b_1+r a_1 b_3,...,...)$$
from which we have the relation
$$x~a_2b_2+y~a_3b_3+z~a_1b_1+t~a_3b_1+r a_1 b_3=\tilde{x}~a_1b_2+
\tilde{y}~a_1b_3+\tilde{z}~a_2b_1+\tilde{t}~a_3b_1 $$
from which we get
$$x=y=\tilde{x}=\tilde{z}=z=0,~~t=\tilde{t},~~r=\tilde{y} $$
The final result is 
\begin{eqnarray*}
3=(a_2 b_3,~a_3b_1,~a_1 b_2)\\
\quad\mbox{and}\quad\\
3=(a_3 b_2,~a_1b_3,~a_2 b_1)
\end{eqnarray*}
where the first line comes from terms proportional to $t$ while the second line is proportional
to $r$.
In summary,  with this notation, if $v=(v_1,v_2,v_3)$ is an additional
 triplet, the product of the three 
triplet $a,b$ and $v$ that transform as a singlet 1 in $A_4$ is given by
\begin{equation}\label{eq:abv}   
h_1 (a_2 b_3 v_1+a_3b_1 v_2+a_1 b_2 v_3) + h_2  (a_3 b_2v_1+~a_1b_3v_2+a_2 b_1v_3)
\end{equation}
where $h_1$ and $h_2$ are arbitrary parameters. The term in eq.(\ref{eq:abv}) is invariant 
under $A_4$.

\end{document}